\documentclass[prstab,twocolumn,showpacs,preprintnumbers,amsmath,amssymb,superscriptaddress,floatfix]{revtex4}
\usepackage{graphicx}

\begin{document}
\title{High precision beam momentum determination in a synchrotron using a spin-resonance method}

\author{P.~Goslawski}
\email{paul.goslawski@uni-muenster.de} \affiliation{Institut f\"ur
Kernphysik, Universit\"at M\"unster, D-48149 M\"unster, Germany}
\author{A.~Khoukaz}
\affiliation{Institut f\"ur Kernphysik, Universit\"at M\"unster,
D-48149 M\"unster, Germany}
\author{R.~Gebel}
\affiliation{Institut f\"ur Kernphysik and J\"ulich Centre for Hadron
Physics, Forschungszentrum J\"ulich, D-52425 J\"ulich, Germany}
\author{M.~Hartmann}
\affiliation{Institut f\"ur Kernphysik and J\"ulich Centre for Hadron
Physics, Forschungszentrum J\"ulich, D-52425 J\"ulich, Germany}
\author{A.~Kacharava}
\affiliation{Institut f\"ur Kernphysik and J\"ulich Centre for Hadron
Physics, Forschungszentrum J\"ulich, D-52425 J\"ulich, Germany}
\author{A.~Lehrach}
\affiliation{Institut f\"ur Kernphysik and J\"ulich Centre for Hadron
Physics, Forschungszentrum J\"ulich, D-52425 J\"ulich, Germany}
\author{B.~Lorentz}
\affiliation{Institut f\"ur Kernphysik and J\"ulich Centre for Hadron
Physics, Forschungszentrum J\"ulich, D-52425 J\"ulich, Germany}
\author{R.~Maier}
\affiliation{Institut f\"ur Kernphysik and J\"ulich Centre for Hadron
Physics, Forschungszentrum J\"ulich, D-52425 J\"ulich, Germany}
\author{M.~Mielke}
\affiliation{Institut f\"ur Kernphysik, Universit\"at M\"unster,
D-48149 M\"unster, Germany}
\author{M.~Papenbrock}
\affiliation{Institut f\"ur Kernphysik, Universit\"at M\"unster,
D-48149 M\"unster, Germany}
\author{D.~Prasuhn}
\affiliation{Institut f\"ur Kernphysik and J\"ulich Centre for Hadron
Physics, Forschungszentrum J\"ulich, D-52425 J\"ulich, Germany}
\author{R.~Stassen}
\affiliation{Institut f\"ur Kernphysik and J\"ulich Centre for Hadron
Physics, Forschungszentrum J\"ulich, D-52425 J\"ulich, Germany}
\author{H.J.~Stein}
\affiliation{Institut f\"ur Kernphysik and J\"ulich Centre for Hadron
Physics, Forschungszentrum J\"ulich, D-52425 J\"ulich, Germany}
\author{H.~Stockhorst}
\affiliation{Institut f\"ur Kernphysik and J\"ulich Centre for Hadron
Physics, Forschungszentrum J\"ulich, D-52425 J\"ulich, Germany}
\author{H.~Str\"oher}
\affiliation{Institut f\"ur Kernphysik and J\"ulich Centre for Hadron
Physics, Forschungszentrum J\"ulich, D-52425 J\"ulich, Germany}
\author{C.~Wilkin}
\affiliation{Physics and Astronomy Department, UCL, London WC1E 6BT,
United Kingdom}

\date{\today}

%
\begin{abstract}
In order to measure the mass of the $\eta$ meson with high accuracy
using the $dp\to\,^3\textrm{He}\,\eta$ reaction, the momentum of the
circulating deuteron beam in the Cooler Synchrotron COSY of the
Forschungszentrum J\"ulich has to be determined with unprecedented
precision. This has been achieved by studying the spin dynamics of
the polarized deuteron beam. By depolarizing the beam through the use
of an artificially induced spin resonance, it was possible to
evaluate its momentum $p$ with a precision of $\Delta p/p < 10^{-4}$
for a momentum of roughly 3~GeV/$c$. Different possible sources of
error in the application of the spin-resonance method are discussed
in detail and its possible use during a standard experiment is
considered.
\end{abstract}

\maketitle

%
\section{Introduction}
\label{sec:intro} For numerous high precision experiments, knowing
the beam momentum in an accelerator with the greatest accuracy is
essential. Obvious examples of this are investigations of production
reactions very close to the thresholds as well as particle mass
determinations on the basis of reaction kinematics. Here we present a
technique that allows one to determine the momentum of a deuteron
beam which is suitable for use in a precise measurement of the mass
of the $\eta$ meson.

Measurements of the mass of the $\eta$ meson performed at different
experimental facilities over the last decade have resulted in very
precise results which differ by up to 0.5~MeV/$c^2$, i.e., by more
than eight standard deviations. The experiments that are no longer
considered in the PDG tables~\cite{pdg2008} generally involve the
identification of the $\eta$ as a missing-mass peak produced in a
hadronic reaction. In order to see whether this is an intrinsic
problem, and to clarify the situation more generally, a refined
measurement of the $dp\to\mathrm{^3He}\,\eta$ reaction was
proposed~\cite{khoukaz2007} at the Cooler Synchrotron COSY of the
Forschungszentrum J\"{u}lich~\cite{maier1997}.

After producing the $\eta$ mesons through the
$dp\to\mathrm{^3He}\,\eta$ reaction using a hydrogen cluster-jet
target~\cite{khoukaz1999}, the $^3$He would be detected with the ANKE
magnetic spectrometer~\cite{barsov2001} that is located at an
internal-target position of the storage ring. Provided that the
reaction is cleanly isolated, the $\eta$ mass can be extracted from
pure kinematics through the determination of the production
threshold. This requires one both to identify the threshold and to
measure accurately the associated beam momentum.

We have previously proved that ANKE has essentially 100\% acceptance
for the $dp\to\mathrm{^3He}\,\eta$ reaction for excess energies $Q$
below about 10~MeV~\cite{mersmann2007}, though in that experiment the
deuteron beam was continuously ramped from below the threshold up to
$Q\approx 11$~MeV. However, although the threshold was well
identified, the corresponding value of the beam momentum was only
known in the experiment with a relative accuracy of about $10^{-3}$.

For the new $\eta$ mass proposal~\cite{khoukaz2007}, the decision was
taken to measure at thirteen fixed energies in the range $1<Q<10$~MeV
as well as $Q = -5$~MeV for background studies. To determine the mass
using this kinematic method with a precision that is competitive with
other recent measurements, i.e., $\Delta
m_{\eta}<50$~keV/$c^2$~\cite{pdg2008}, the associated beam momenta
have to be fixed with an accuracy of $\Delta p /p < 10^{-4}$. This
requires the thirteen beam momenta in the range of  $3100 -
3200$~MeV/$c$ to be measured to better than 300~keV/$c$.

Generally at synchrotron facilities like COSY, the velocity of the
beam particles, and hence the beam momentum, is determined from the
knowledge of the revolution frequency combined with the absolute
orbit length. The accuracy that can be reached using this technique
is limited by the measurement of the orbit length by, e.g., beam
position monitors. This is in the region of $\Delta p/p \approx
10^{-3}$ and so an order of magnitude improvement is needed for the
$\eta$ mass experiment. Because of the technical limitations of such
a macroscopic device, it is not feasible to obtain the necessary
increase in accuracy by simply scaling up the number of beam pick-up
electrodes. The beam momentum must therefore be determined in some
other way.

The method proposed for electron colliders more than thirty years ago
to overcome this problem~\cite{Serednyakov76,Derbenev1980} has been
very successfully applied at the VEPP accelerator of the BINP at
Novosibirsk to measure the masses of a wide variety of mesons from
the $\phi$ to the $\Upsilon$~\cite{VEPP}. The technique was further
developed at DORIS in Hamburg~\cite{DORIS} and CESR in
Cornell~\cite{CESR} as well as LEP at CERN~\cite{LEP}.

The spin of a polarized beam particle precesses around the normal to
the plane of the machine, which is generally horizontal. The spin can
be perturbed by the application of a horizontal \emph{rf} magnetic
field from, for example, a solenoid. The beam depolarizes when the
frequency of the externally applied field coincides with that of the
spin precession in the ring. The usefulness of the technique relies
on the fact that a frequency $f$ can be routinely measured with a
relative precision of $\Delta f/f = 10^{-5}$. Furthermore, the
position of the depolarizing resonance depends purely upon the
revolution frequency of the machine and the kinematical factor
$\gamma=E/mc^2$, where $E$ and $m$ are the particle total energy and
mass, respectively. The measurements of the revolution and
depolarizing frequencies together allow the evaluation of $\gamma$
and hence $E$ and the beam momentum $p$.

There is no in-principle reason why the induced-depolarization
approach should not be equally applicable to other beam particles
with an intrinsic spin, such as protons or deuterons. In fact, the
effects have recently been confirmed at COSY in studies of the spin
manipulation of both polarized proton~\cite{morozov2004} and deuteron
beams~\cite{morozov2005}. This is the methodology that we are
pursuing at COSY for the measurement of the $\eta$ mass. For the
first time in 2007 it was possible in a test run to reach an accuracy
in the beam momentum calibration of $\Delta p / p < 10^{-4}$ using
the technique with a coasting beam but no internal
target~\cite{stockhorst2007}. In the present paper we describe how
the method can be used in a standard beam time under normal
experimental conditions in the presence of a thick internal target.

In Sec.~\ref{sec:theory} we describe the physical principles
underlying the spin-resonance method. After discussing the behavior
of a vector polarized deuteron beam in COSY, we show how to induce an
artificial spin resonance to depolarize the beam. The experimental
conditions that allow one to determine the two critical observables
are explained in Sec.~\ref{sec:conditions}. The revolution frequency
$f_{0}$ is measured via the Schottky noise of the beam and the
spin-resonance frequency $f_{r}$ using the \emph{rf} solenoid and the
EDDA detector as a beam polarimeter~\cite{EDDA}. The deuteron beam
results are presented in Sec.~\ref{sec:results}, where the estimated
uncertainties are discussed in some detail. Our conclusions are
summarized in Sec.~\ref{sec:end}.

%
\section{Theoretical background of the spin-resonance method}
\label{sec:theory}
%
\subsection{Spin in synchrotrons}
In contrast to the case of a spin-half fermion such as an electron or
proton, the deuteron is a spin-one boson that can be placed in three
magnetic sub-states $m~=~{-1},\: 0,\:{+1}$, and the resulting
polarization phenomenology is more complex. Eight independent
parameters are necessary to characterize a spin-one beam, three for
the vector polarization and five for the tensor~\cite{Ohlsen}.
However, only the vector polarization
\begin{equation}
    P_{V} = (N_{+} - N_{-}) / N\,,
\end{equation}
is used in the present experiment for the spin-resonance method since
it can be measured with the beam polarimeter to a higher precision
than the tensor. Here $N_m$ is the number of particles in state-$m$
and $N = N_{+} + N_{-} + N_{0}$ is the total number of particles.

The motion of the spin vector $\vec{S}$, defined in the rest frame of
the particle, in a circular accelerator, synchrotron or storage ring,
is given by the Thomas-BMT equation~\cite{BMT}:
\begin{widetext}
  \begin{equation}
  \label{eq:BMT}
    \frac{d\vec{S}}{dt} = \frac{e}{\gamma m}\vec{S}\times
    \left[\left(1+\gamma G\right)\vec{B}_{\bot} +
    \left(1+ G \right)\vec{B}_{||} +
    \left( G\gamma + \frac{\gamma}{\gamma + 1} \right) \frac{\vec{E} \times \vec{\beta} }{c}\right]  \; ,
  \end{equation}
\end{widetext}
where $\vec{B}_{\bot}$ and  $\vec{B}_{||}$ are the transverse and
longitudinal components of the magnetic fields of the accelerator in
the laboratory frame and $\vec{E}$ represents the electric field. The
velocity of the particle is $\vec{\beta} c$, in terms of which
$\gamma=1/\sqrt{1-\beta^2}$.

In a synchrotron without horizontal magnetic fields and where the
electric field is always parallel to the particle motion, the spin
motion only depends on the first term, i.e., is a function of the
transverse magnetic fields~$\vec{B}_{\bot}$ of the accelerator. The
deuteron spin precesses around the stable spin direction, which is
given by the vertical fields of the guiding dipole magnets of the
synchrotron. The number of spin precessions during a single circuit
of the machine, the spin tune $\nu_{s}$, is proportional to the
particle energy. In the coordinate basis of the moving particle, the
spin tune is given by
\begin{equation}
\label{eq:SpinTune}
    \nu_{s} = G \,\gamma\, ,
\end{equation}
whereas, taking into account the extra rotation associated with a
single circuit of the machine, this becomes $\nu_{s} = 1+G \, \gamma$
in the laboratory frame. Here $G=(g-2)/2$ is the gyromagnetic anomaly
of the particle, where $g$ is the gyromagnetic factor. For deuterons
the gyromagnetic anomaly, $G_d=-0.1429872725\pm0.0000000073$, can be
calculated from the ratios of the magnetic moments and masses of the
proton and deuteron~\footnote{Generally the $g$-factor of the
deuteron is written in units of the nuclear magneton. If it is
required in terms of the deuteron magneton, this has to be calculated
from the $g$-factor of the proton and the ratios of the magnetic
moment and mass of proton and deuteron: $g_d=\frac{1}{2}
g_p\mu_dm_d/\mu_pm_p$. The constants required can be found in the
NIST compilation~\protect{\cite{nist}} but in the evaluation of the
uncertainty one has taken into account the fact that the values
$(\mu_d/\mu_p\,,g_p)$ and $(m_d/m_p\,,g_p)$ are correlated. Having
done this, we find $G_d = -0.1429872725 \pm 0.0000000073$.}.
%
\subsection{Artificially induced depolarizing resonances}
The beam polarization can be perturbed by a horizontal magnetic field
in the synchrotron and, if the frequency of the perturbation
coincides with the spin-precession frequency, the beam depolarizes.
One kind of first-order resonance is the imperfection resonance. If
the spin tune is an integer, then the horizontal imperfection fields
of the synchrotron can interact resonantly with the particle spin,
building up effects coherently turn by turn. The positions in
momentum of the depolarizing resonances depend on the gyromagnetic
anomaly of the particle. In contrast to the case of protons, where
the first imperfection resonance occurs at a momentum of
$464$~MeV/$c$, the first for deuterons is at $13$~GeV/$c$, which is
well outside the COSY momentum range. Furthermore, in the present
experiment the spin tune remains in the region of $\nu_s = 0.2775 -
0.2818$.

Because of the betatron oscillation frequency of the circulating
beam, the particles can also encounter the fields of the focusing
quadrupole magnets in resonance with the spin tune, which lead to a
second kind of first-order spin resonance, the so-called intrinsic
resonance. These resonances also occur only for energies that are far
beyond the COSY deuteron momentum and working point
range~\cite{lehrach2003}.

A horizontal \emph{rf} field from a solenoid can lead to
\emph{rf}-induced depolarizing resonances. Depending on the form of
the field, these can be used to depolarize the beam, to measure the
spin tune, or even to flip the spin direction of the beam particles.
The spin-resonance frequency for a planar accelerator where there are
no horizontal fields is given by~\cite{Derbenev1980}
\begin{equation}
\label{eq:SpinRes}
    f_{r} = (k + \gamma \, G) f_{0} \;,
\end{equation}
where $f_0$ is the revolution frequency of the beam, $\gamma \, G$ is
the spin tune, and $k$ is an integer. If the \emph{rf} frequency of
the perturbation is close to $f_r$ then the polarization of the beam
is maximally influenced. Horizontal magnetic fields in the
accelerator lead to modifications of Eqs.~\eqref{eq:SpinTune} and
\eqref{eq:SpinRes}~\cite{LEP, SYLEE}. To avoid this complication, all
solenoidal and toroidal magnets in the COSY ring, those of the
experiment as well as those of the electron cooler, were switched
off. Residual shifts in the resonance frequency arising from field
errors and vertical orbit distortions were estimated and found to be
negligibly small. These effects are discussed in more detail in
Sec.~\ref{subsec:ResSysShift}.

It is important to note that Eq.~\eqref{eq:SpinRes} is only valid if,
as is the case for the present experiment, there is no full or
partial Siberian snake. The resonance with $k=1$ was used as this
matches the frequency range of the \emph{rf} solenoid installed at
COSY. The kinematic $\gamma$-factor, and thus the beam momentum, can
be determined purely by measuring both the revolution and
spin-resonance frequencies.
%
\section{Experimental conditions}
\label{sec:conditions} 
\begin{figure}[hbt]
\includegraphics[width=1.0\linewidth]{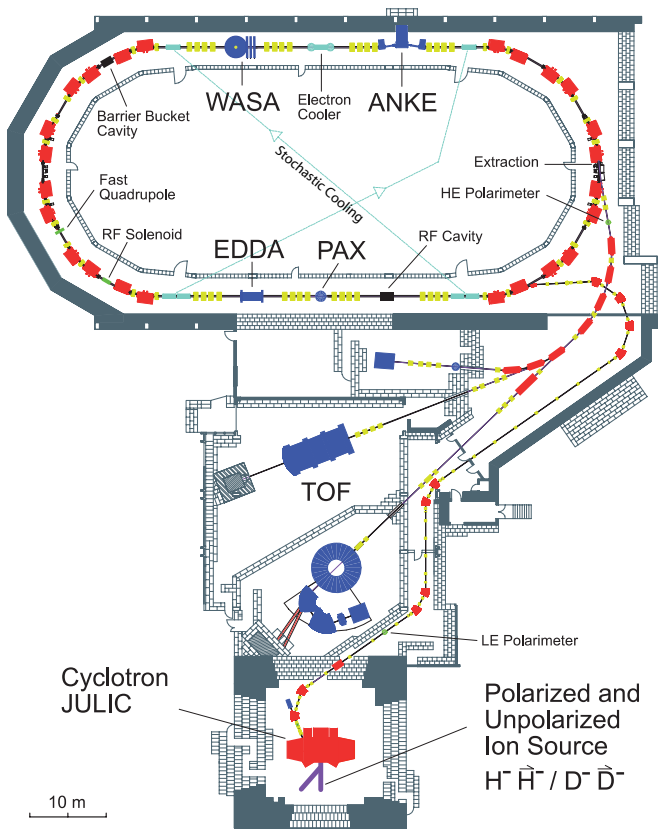}
\caption{\label{fig:COSY} (Color online) The COSY accelerator
facility. The cyclotron JULIC provides both unpolarized and polarized
proton and deuteron beams for injection into the COSY ring, where
they are accelerated and stored. COSY operates in the momentum range
of $300-3700$~MeV/$c$. The position of the ANKE spectrometer with the
thick internal hydrogen cluster-jet target is shown, as are those of
the \emph{rf} solenoid to depolarize the deuteron beam, the barrier
bucket cavity to compensate beam-target energy losses, and the EDDA
detector that was used as a beam polarimeter.}
\end{figure}

The COSY accelerator facility is presented in
Fig.~\ref{fig:COSY}. After pre-acceleration in the Cyclotron JULIC,
COSY can provide unpolarized and polarized proton and deuteron beams
in the momentum range of $300-3700$~MeV/$c$. For the present
experiment, two of the four internal facilities were used, viz.\ ANKE
with a thick hydrogen cluster-jet target and EDDA~\cite{EDDA} as beam
polarimeter~\cite{CHI08}. The beam was accelerated with the \emph{rf}
cavity and the barrier bucket (\emph{bb}) cavity was used to
compensate for the energy losses incurred through the beam-target
interactions (see Sec.~\ref{subsec:cavity}). The position of the
\emph{rf} solenoid is also shown. The integrated value of the
solenoid's maximum longitudinal \emph{rf} magnetic field is $\int
B_{\text{rms}}\;dl=0.67$~T~mm at a \emph{rf} voltage of 5.7~kV rms.
Its frequency range is 0.5--1.5~MHz.

%
\subsection{The \emph{rf} cavity system}
\label{subsec:cavity}%
For a high precision experiment it is crucial that the beam momentum
remains stable throughout the whole accelerator cycle. In a typical
cycle of a standard scattering experiment at ANKE, the beam is first
injected into COSY and accelerated to the nominal momentum. The
\emph{rf} cavity is then switched off to provide a coasting beam that
fills the ring uniformly. This then gives constant count rates, which
reduces the dead time of the data acquisition system (DAQ). But,
because of the energy losses of the charged beam particles through
electromagnetic processes as the beam passes repeatedly through the
target, the momentum changes and this leads to a shift in the
revolution frequency~\cite{stein2008}. For a deuteron beam and a
hydrogen cluster-jet target with a density of $\rho = 1 \times
10^{15}~\mathrm{cm}^{-2}$, the revolution frequency would change by
up to $103 \ \mathrm{Hz}$ over a 180~s long cycle, corresponding to a
shift in beam momentum of 2.2~MeV/$c$.

To compensate for this effect and to guarantee a constant beam
momentum over the whole data-taking cycle, a second cavity, the
barrier bucket (\emph{bb}) cavity~\cite{stassen2008}, was switched on
after the \emph{rf} cavity was switched off. In this way a beam with
a constant momentum over the whole cycle could be produced that
filled roughly 80 -- 90\% of the ring homogenously and thus achieved
the necessary reduction in the dead time of the DAQ.
%
\subsection{The cycle timing and the supercycle}
\label{subsec:cycletiming}%
The thirteen closely spaced energies studied near the $\eta$
threshold were divided into two so-called supercycles that involved
up to eight different COSY machine settings. The first and the second
supercycle each consisted of seven different energies where, to allow
comparison between the two sets, the first energies of the two
supercycles were chosen to be identical. Data at thirteen different
energies were therefore recorded. The different machine settings in
the supercycles were imposed sequentially, after which the supercycle
was repeated. Each supercycle was used for five days of continuous
Schottky data taking to study the long term stability of COSY and to
take data in parallel for the $\eta$ meson mass determination. The
reason for choosing supercycles instead of independent measurements
at fixed energies was to guarantee the same experimental conditions
for each of the beam energies in one supercycle. In this way the systematic
uncertainties could be investigated in more detail, as will be
discussed in Sec.~\ref{sec:results}.

\begin{table}[hbt]
\caption{\label{tab:TimingSRM} Cycle timings used to determine the
spin-resonance frequency spectrum with the polarized beam.}
    \begin{ruledtabular}
        \begin{tabular}{l l}
            Time (s)        &  Process\\
            \hline
            0                   &  Start of cycle: injection\\
            0 -- 3.7            &  Acceleration of the beam with \emph{rf} cavity\\
            3.7                 &  Switch off \emph{rf} cavity\\
            4                   &  Switch on \emph{bb} cavity\\
            20 -- 25        &  \emph{rf} solenoid on\\
            25 -- 30            &  Polarization measurement with EDDA\\
            36                  &   End of cycle\\
        \end{tabular}
    \end{ruledtabular}
\end{table}

Before starting each of the five day blocks, the individual beam
energies were measured using 36~s accelerator cycle lengths. The
timing structure of the accelerator cycles is described in
Table~\ref{tab:TimingSRM}. After the injection of the beam into COSY,
the stored deuterons were accelerated to the first nominal beam
energy of the supercycle using the regular COSY \emph{rf} cavity. At
$t=3.7$~s this cavity was switched off and at $t = 4$~s the \emph{bb}
cavity was brought into operation to compensate for the beam energy
losses. At $t=20$~s the amplitude of the depolarizing \emph{rf}
solenoid was linearly ramped from 0 to 2.4~kV rms to produce a $\int
B_{\text{rms}} \; dl = 0.29$~T mm in 200~ms, remained constant for
5~s, and was then ramped down in 200~ms. This was followed by a beam
polarization measurement for five seconds using the EDDA
detector~\cite{EDDA}. At $t = 36$~s the cycle was terminated. This
procedure was repeated at the same beam energy but with different
\emph{rf} solenoid frequencies in order to obtain the spin-resonance
spectrum. After completion of this first sub-measurement, the next
beam energy of the supercycle was used and the corresponding
spin-resonance spectrum measured until complete data was obtained at
all the energies of the supercycle.

After measuring the spin-resonance spectrum, the supercycle was
switched on for five days of continuous data taking to investigate
the long term stability of the COSY accelerator. For this study the
polarization measurements were omitted and total cycle lengths of
206~s were used. After injection, acceleration and starting the
\emph{bb} cavity, Schottky measurements were performed over the time
interval of $t=14-196$~s. The eight beam energies in one supercycle (the first one is installed twice)
involved a total time of 1648~s, after which the supercycle was
repeated. After the five days of data taking, the system was returned
to the conditions of Table~\ref{tab:TimingSRM} to repeat the
measurement of the spin-resonance spectrum in order to control
systematic effects.

The polarized ion source at the injector cyclotron of COSY currently
gives a beam intensity that is about an order of magnitude too low
compared to that which is required for the $\eta$ mass proposal. It
was therefore decided to use this ion source only  for the beam
energy measurement before and after the supercycles. As a
consequence, for the long term stability studies COSY was switched to
the unpolarized ion source, which allowed beam intensities up to
$n_d\approx 1\times10^{10}$. However, it had to be carefully checked
that the same COSY beam energies were obtained when using the
polarized and the unpolarized ion sources. To ensure this, the
complete settings of the cyclotron, the beam injection, as well as
COSY itself, were fixed when switching from one ion source to the
other. The revolution frequencies $f_0$ of the stored beam in the two
cases matched to within $\Delta f_0\, \le \,6$ Hz, proving the
validity of this method. The determination of $\Delta f_0$ was
limited in the present case by the experimental resolution of the
Schottky spectrum analyzer, though this could be improved by better
calibration.

Both in the beam energy determination, as well as later in the
Schottky data-taking time, one had to be assured that the
measurements within the cycles were started sufficiently long after
the ramping of the COSY dipole magnets for the acceleration of the
beam. Otherwise, the not-yet-stable magnetic fields would lead to
deviations in the values determined for the beam momentum. Detailed
measurements of the beam energy by the spin-resonance method as a
function of the time in the cycle showed that the experimental
situation is already stable ten seconds after the start of the
cycle~\cite{Pauldipl}. Therefore, the \emph{rf} solenoid field and
the Schottky data-taking were started 20 and 14~s after injection,
respectively. As a further check, measurements showed that the same
beam energy was observed close to the end of the cycle as at the
beginning~\cite{Pauldipl} (see Sec.~\ref{subsec:Resfr}). Thus it is
valid to investigate the beam energy at one fixed time during the
cycle and to take the resulting value as representative for the whole
cycle.
%
\subsection{Determination of the revolution frequency $\boldsymbol{f_{0}}$
via the Schottky noise measurements} \label{subsec:f0}

The revolution frequency $f_0$ was measured by using the Schottky
noise of the deuteron beam. The origin of this effect is the
statistical distribution of the charged particles in the beam. This
leads to random current fluctuations that induce a voltage signal at
a beam pick-up in the ring. The Fourier transform of this
voltage-to-time signal by a spectrum analyzer delivers the frequency
distribution around the harmonics of the revolution frequency of the
beam.

For the measurement of the Schottky noise, the beam pick up and the
spectrum analyzer of the stochastic cooling system of COSY were used.
The spectrum analyzer (standard swept-type model HP 8595E) is
sensitive to the Schottky noise current, which is proportional to the
square root of the number $N$ of the particles in the beam. To get
the Schottky power spectra, which represent the momentum
distribution~\cite{boussard1987}, the amplitudes of the measured
distribution were squared. The spectrum analyzer was operated in the
range of the thousandth harmonic but, because of the different flat
top revolution frequencies, harmonics from 997 to 1004 were also
measured.

\begin{figure}[htb]
\includegraphics[width=1.0\linewidth]{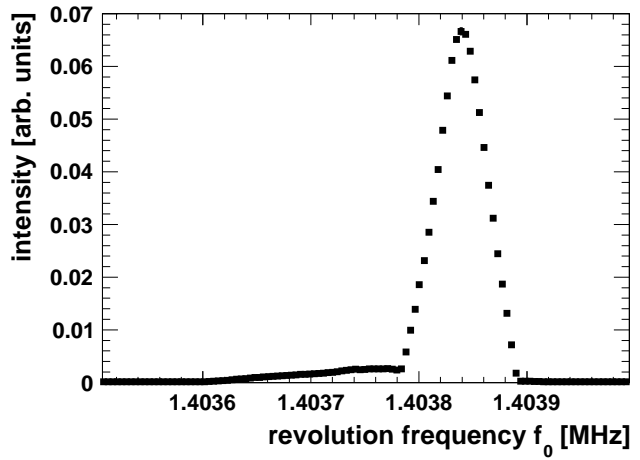}
\caption{\label{fig:f0_SC1FL1} Mean Schottky power
spectrum extracted from measurements over five days at one energy.
The statistical error bars lie within the data points. By calculating
the weighted arithmetic mean, an average revolution frequency of
$\overline{f_0} = 1403831.75 \pm 0.12$~Hz was deduced.}
\end{figure}

The Schottky spectra were recorded every 30~s throughout the whole
beam time so that altogether nearly 15000 distributions were
collected and sorted by energy, i.e., by flat top. This large number
of Schottky measurements allows the study of the long term stability
of the revolution frequency, which will be discussed in
Sec~\ref{sec:results}. From all the spectra taken over five days that
were measured under the same conditions at a particular energy, one
mean spectrum was calculated, an example of which is presented in
Fig.~\ref{fig:f0_SC1FL1}.

The full width at half maximum is in the region of $40-50$~Hz for all
energies. The position of the mean distribution of the circulation
frequency is stable for the whole cycle time but, within the cycle, a
small tail is seen at lower frequencies. This corresponds to beam
particles that escaped the influence of the \emph{bb} cavity but
still circulated in COSY. By calculating the weighted arithmetic mean
of the revolution frequency distribution, an average revolution
frequency was estimated. The statistical uncertainty of the mean
revolution frequency, which is below 0.2~Hz for all energies, depends
on both the number of measured Schottky spectra and on the
distribution variations.
%
\subsection{Determination of the spin-resonance frequency $\boldsymbol{f_{r}}$
via an induced spin resonance} \label{subsec:fr}

For all thirteen energies the spin-resonance spectrum was measured
twice, once before and once after the five days of Schottky data
taking, as described in Sec.~\ref{subsec:cycletiming}. The
polarization of the beam leads to an asymmetry in scattering from a
carbon target, which was measured with the EDDA
detector~\cite{schwarz1999}. For our purposes absolute calibrations
of this device at the different energies were not required; a
quantity merely proportional to the polarization such as the
left-right asymmetry is sufficient.

\begin{figure}[hbt]
\includegraphics[width=1.0\linewidth]{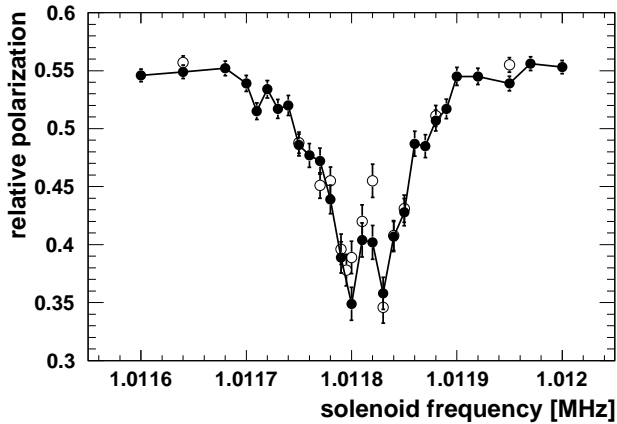}
\caption{\label{fig:fr_BegEnd}Spin-resonance measurements at one
energy (closed circles). The cycle timings are described in
Table~\ref{tab:TimingSRM}. The open symbols represent results
obtained for an extended cycle time, where the perturbing solenoid
was switched on after 178~s.}
\end{figure}

An example of a spin-resonance spectrum at one energy is shown in
Fig.~\ref{fig:fr_BegEnd}. This displays the non-normalized
polarization (``relative polarization'') as a function of the
solenoid frequency. Far away from the spin resonance at 1.0116~MHz
and 1.0120~MHz, a high beam polarization was measured. In contrast,
when the frequency of the solenoid coincided with the spin-precession
frequency, the beam was maximally depolarized. The full width at half
maximum was in the region of 80-100~Hz for all energies. Unlike the
earlier spin-resonance test measurement with a coasting beam, i.e.,
no cavities and no internal target~\cite{stockhorst2007}, the
spin-resonance spectra are not smooth. The structures, especially the
double peak in the center, are caused by the interaction of the
deuteron beam with the \emph{bb} cavity. However, by comparing the
spin-resonance spectra measured for an unbunched and bunched beam
with accelerating cavity with $h = 1$ or the barrier bucket cavity,
it was found that the centers of gravity of the spectra were the
same.

\begin{figure}[htb]
\includegraphics[width=1.0\linewidth]{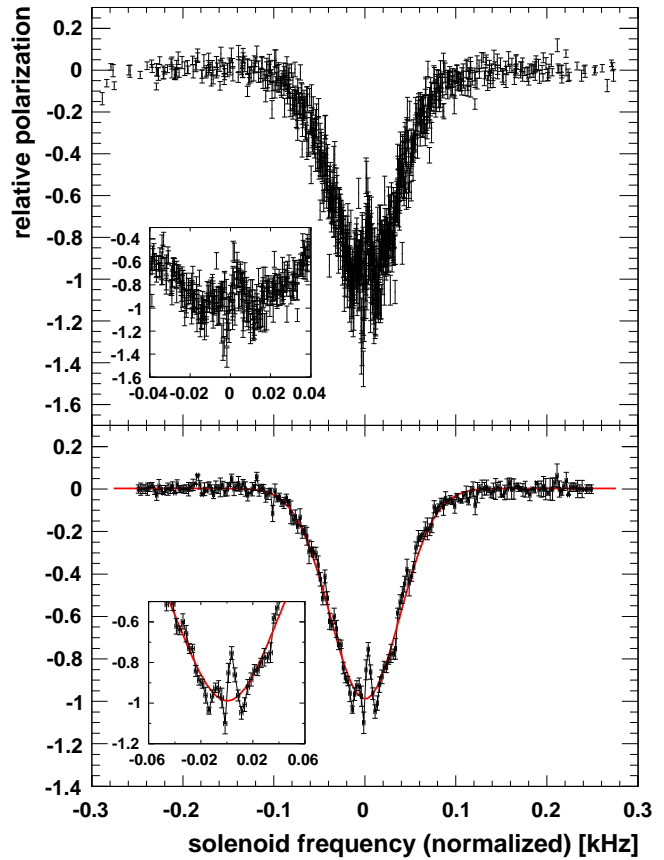}
\caption{\label{fr_BinnedNorm} Panel (a): The spin-resonance spectra
normalized by a gaussian. Panel (b): The same but with larger bins.
The spin-resonance shape is symmetric about zero and smooth except in
the center, where a double peak structure is seen. The structures,
especially the double peak in the center, are caused by the
interaction of the deuteron beam with the \emph{bb} cavity. The
inserts show the resonance valley in greater detail.}
\end{figure}

To study the shapes of the spin-resonance spectra in more detail, all
26 distributions were fitted with gaussians and then shifted along
the abscissa so that the mean value of each individual spectrum was
zero. In addition, each spectrum was shifted along the ordinate so
that the off-resonance polarization vanished. Finally, the data were
scaled to a uniform height and displayed together in a single plot to
allow a comparison of all the spectra. The resulting global
spin-resonance spectrum shown in Fig.~\ref{fr_BinnedNorm}(a) is
symmetric around zero and smooth, except for the structure at the
center. This region is shown in greater detail in the insert. In
order to improve the visibility of the structures close to the
minimum, the size of the frequency bins was increased and the results
displayed in Fig.~\ref{fr_BinnedNorm}(b).

A structure with a symmetric double peak and an oscillation is
observed in the center of the spin resonance. However, it is
important to note that the gaussian mean value, i.e., the
spin-resonance frequency, is not influenced by this structure. This
was checked by making a fit where the data points at the center were
excluded. The spin-resonance frequencies $f_r$ were extracted from
the spin-resonance spectra for all energies by making gaussian fits.
These gave $\chi^2$/ndf in the region of 2--3. The statistical
uncertainties of the spin-resonance frequencies are on the order of
1--2~Hz at $f_r \approx 1.01$~MHz.

%
\section{Results}
\label{sec:results}
%
\subsection{Stability of the revolution frequency $\boldsymbol{f_{0}}$}
\label{subsec:Resf0} The \emph{bb} cavity compensates the effects of
beam-target energy losses and should ensure that the revolution
frequency remains constant. The large number of Schottky measurements
allowed us to study the long term stability and to identify the
magnitude of the variations  of the revolution frequency at COSY.
Therefore all the Schottky spectra at one energy from one day were
analyzed and the mean revolution frequency of that day calculated, as
described in Sec~\ref{subsec:f0}.

\begin{figure}[hbt]
\includegraphics[width=1.0\linewidth]{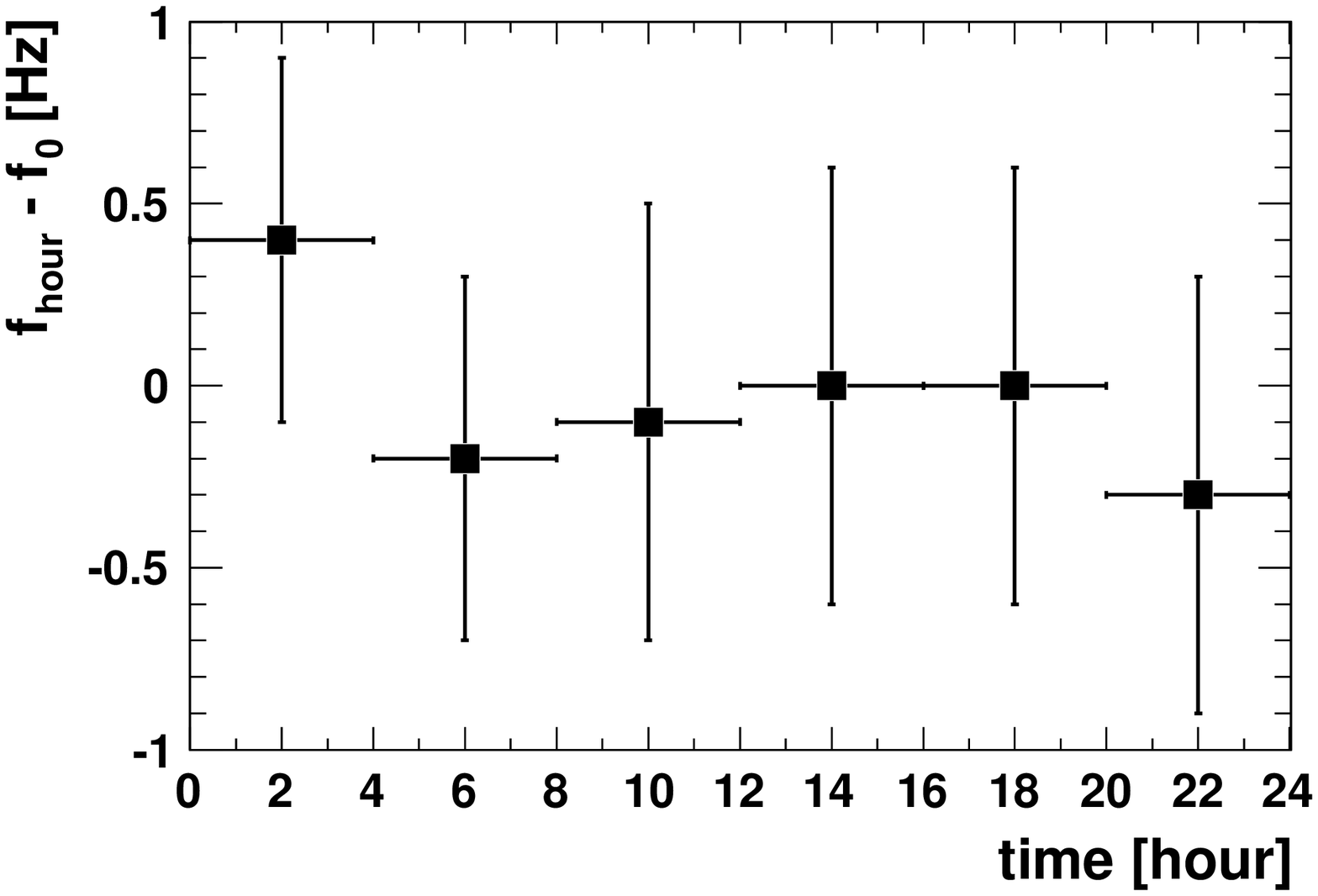} \\[0.2cm]
\includegraphics[width=1.0\linewidth]{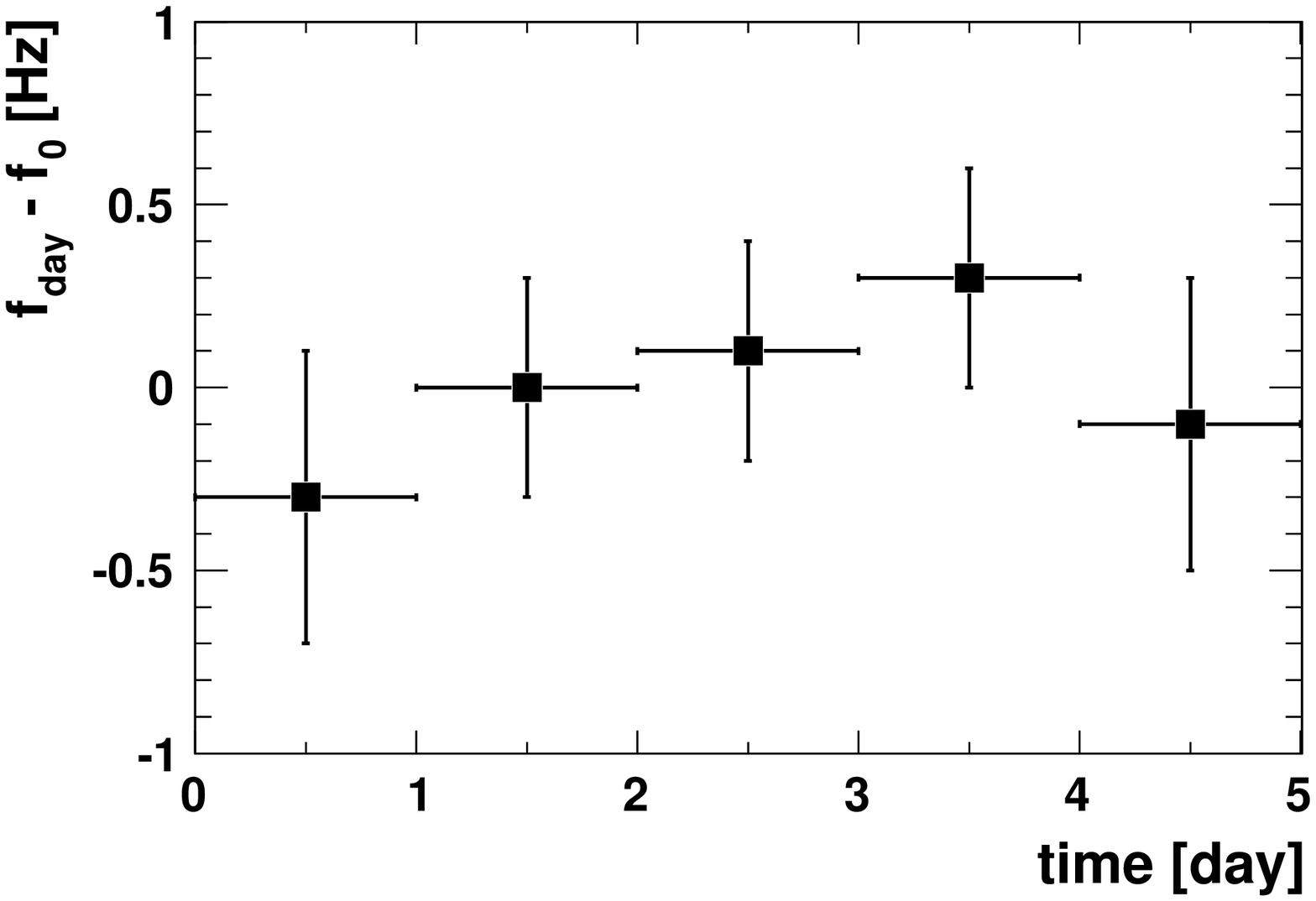}
\caption{\label{fig:f0_Stab} Stability of the revolution frequency
$f_0$. In panel (a) the differences between the revolution
frequencies for every four hours and the mean revolution frequency of
the day are shown. In panel (b) the differences between the
revolution frequencies for each day and the mean revolution frequency
of the five days of Schottky data taking are shown. From these
figures it is clear that the revolution frequency at COSY is very
stable, with variations below $1$~Hz at a circulation frequency of
$f_0 \approx 1.4$~MHz.}
\end{figure}

In addition, the revolution frequencies for these data were
calculated for every four hours to study the daily variation of the
circulation frequency. The differences between the revolution
frequencies of every four hours and the mean frequency of the day are
presented in the upper part of Fig.~\ref{fig:f0_Stab}. To study the
variation of the revolution frequency over the five days of data
taking, the same procedure was carried out for the Schottky data
measured over this period. The differences between the mean
revolution frequencies of every day and the mean frequency of the
whole five days of data taking are presented in the bottom part of
Fig.~\ref{fig:f0_Stab}. The horizontal bars represent the time
intervals for which the revolution frequency was evaluated.

The analysis shows that the revolution frequency at COSY over one day
and also over five days is very stable. The variations of the
revolution frequency are very small, being on the order of $1$~Hz at
a circulation frequency of $f_0 \approx 1.4$~MHz. In sum, it was
possible to determine the revolution frequencies for all energies
with a statistical uncertainty below $1$~Hz. Nevertheless the much
larger systematic uncertainty of $\Delta f_0 = 6$~Hz  dominated the
precision, and this arose from the preparation of the Schottky
spectrum analyzer used. A more refined calibration of this device
could improve the systematic precision of the circulation frequency
measurement down to 1~Hz.
%
\subsection{Spin-resonance frequency $\boldsymbol{f_{r}}$}
\label{subsec:Resfr} It is important for the interpretation of the
spin-resonance measurements to know to what extent the positions of
the observed spin-resonance frequencies are stable over the finite
accelerator cycle in the presence of a thick internal target.
Therefore, in a special measurement, the switch-on of the \emph{rf}
solenoid was delayed from 20~s to 178~s in order to investigate the
position of the spin-resonance frequency close to the end of a long
cycle. The observed data (open symbols of Fig.~\ref{fig:fr_BegEnd})
showed a resonance position which agreed with the data taken at the
beginning of the cycle to within 2~Hz.

\begin{figure}[hbt]
\includegraphics[width=1.0\linewidth]{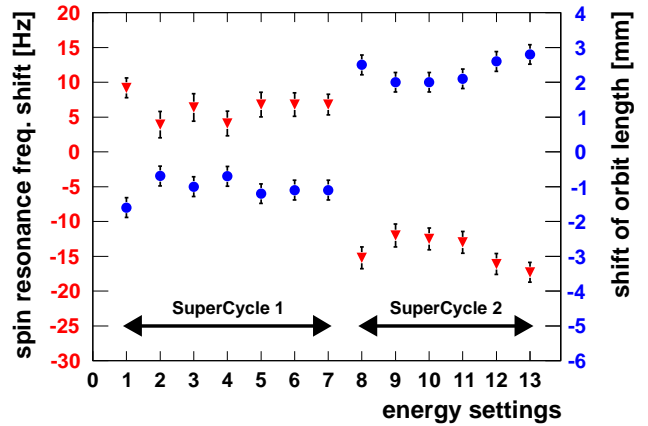}
\caption{\label{fig:frol} (Color online) The spin-resonance
frequencies were measured twice, once before and once after the five
days of data taking. The red triangles present the shift of the
spin-resonance frequency $f_r$ from the first to the second
measurement. These shifts correspond to changes in the orbit length,
which are shown as blue circles. For the first supercycle, the
spin-resonance frequencies decrease between the two measurements by
$4-10$~Hz, which corresponds to a increase in the orbit length in the
range of $0.7-1.6$~mm. For the second supercycle an increase of the
spin resonance in the range of $12-17$~Hz was observed, i.e., a
decrease in the orbit length in the range of $2.0-2.8$~mm.}
\end{figure}

In Fig.~\ref{fig:frol} the shifts between the first and second
spin-resonance measurements are shown as red triangles for all
thirteen energies. The frequencies in the first supercycle decrease
by between 4 and 10~Hz for all energies, whereas for the second
supercycle they increase in the range of $12-17$~Hz. These systematic
shifts of the frequencies in the same direction indicate slight
changes in the COSY settings. Because the revolution frequency is
stable, as described in Sec.~\ref{subsec:Resf0}, the change is
attributed to a shift in the orbit length $s$.

The velocity $v$ of the particle is the product of the revolution
frequency and the orbit length $v = s\,f_0$. Using
Eq.~\eqref{eq:SpinRes}, the orbit length can be calculated from the
revolution and the spin-resonance frequencies:
\begin{equation}
s = c\left[\frac{1}{{f_{0}}^2}-\left(\frac{G_{d}}{f_{r} - f_{0}}
\right)^2 \right]^{\frac{1}{2}},
\end{equation}
which allows the orbit lengths to be extracted with a precision
better than $0.3$~mm for every flat top. Since the nominal COSY
circumference is $183.4$~m, this gives a relative accuracy of $\Delta
s / s \leqslant 2 \times 10^{-6}$. The uncertainty is dominated by
that of the spin-resonance frequency. The shift in the spin-resonance
frequency corresponds to a change in the orbit length of up to
$3$~mm, which is presented for all energies in Fig.~\ref{fig:frol} as
blue circles. The shifts of the spin-resonance frequencies of the
first supercycle suggest an increase in the orbit length in the range
of $0.7-1.6$~mm and to a decrease in the range of $2.0-2.8$~mm for
the second supercycle.

To determine the precise beam momenta, the mean value of the two
spin-resonance measurements for every energy was calculated. These
mean values differ by up to 10~Hz from the single spin-resonance
measurements. Nevertheless, in view of the observed shift of the
spin-resonance frequency, a very conservative systematic uncertainty
of $\Delta f_r = 15$~Hz was assumed.
%
\subsection{Accuracy and systematic shifts of the resonance frequency}
\label{subsec:ResSysShift}%

One obvious limitation on the spin-resonance method is given by the
uncertainty in the deuteron gyromagnetic anomaly $G_d$. However, this
leads to a relative precision in the beam momentum of $\Delta p/p =
5\times 10^{-8}$, which can be safely neglected.

The first order uncertainties in the momentum measurement depend on
the accuracies to which the spin-resonance and revolution frequencies
are determined. As described in Sec.~\ref{subsec:Resfr} and
Sec.~\ref{subsec:Resf0}, these are $15~\text{Hz}/1.01~\text{MHz}=1.5
\times 10^{-5}$ and $6.0~\text{Hz}/1.40~\text{MHz}=4.3 \times
10^{-6}$, respectively. The error therefore arises primarily from the
measurement of the spin-resonance frequency.

The intrinsic width of the spin-resonance may also impose a limit on
the accuracy achievable. In this experiment, the integrated value of
the solenoid's maximum longitudinal \emph{rf} magnetic field gives a
resonance strength of about $\epsilon=3\times 10^{-6}$, which leads
to a spin resonance with a FWHM width $\approx 9$~Hz. This is much
smaller than the observed width of 80-100~Hz, which is therefore
dominated by the momentum spread of the beam. Higher order
contributions lead to an additional spread in the spin frequencies
caused by nonlinear synchrotron and betatron
motion~\cite{Lysenko1986}. It should be stressed that these higher
order effects, which are negligible compared to the calculated
resonance width, do not contribute to a shift of the resonances
frequency.

Systematic shifts of the resonance frequencies may be caused by
deviations from idealized conditions in a real accelerator like COSY.
The possible effects and their contribution to the accuracy of the
resonance frequency determination were estimated and are summarized
in Table~\ref{tab:SysShift}.

\begin{table}[hbt]
    \caption{\label{tab:SysShift} Accuracy and possible systematic shifts of the resonance frequency $f_r$.}
    \begin{ruledtabular}
        \begin{tabular}{l | c}
            Source                                                                                                  &       $\Delta f_r/f_r$                    \\
            \hline
            Resonance frequency accuracy from\\  depolarization spectra             &       $1.5 \times 10^{-5}$            \\
            \hline
            Spin tune shifts from longitudinal fields\\ (field errors)                          &       $1.4 \times 10^{-9}$            \\
            \hline
            Spin tune shifts from radial fields\\ (field errors, vertical correctors)       &       $6.0 \times 10^{-9}$            \\
            \hline
            Spin tune shifts from radial fields\\ (vertical orbit in quadrupoles)            &       $4.1 \times 10^{-8}$            
        \end{tabular}
    \end{ruledtabular}
\end{table}

Radial and longitudinal fields in the accelerator may lead to a
modification of Eq.~\eqref{eq:SpinRes}~\cite{SYLEE}, i.e., to a
systematic shift of the resonance frequency. Even though all
solenoidal and toroidal fields, which may act as partial Siberian
snakes, were turned off for this experiment, field errors and
vertical orbit distortions could generate some net radial or
longitudinal fields \cite{LEP, VEPP}. These effects were estimated
for the current experimental conditions and found to be negligibly
small. The typical field errors of the main magnets, $\Delta B/B
\approx 2\times 10^{-4}$, would lead to a shift in the spin-resonance
frequency of $\Delta f_r/f_r < 1.4 \times 10^{-9}$. Similarly, the
observed vertical orbit displacement of $\Delta y_{\text{rms}} <
1.8$~mm would induce a shift of $\Delta f_r/f_r < 6.0 \times
10^{-9}$.

The largest contribution to a systematic shift of the resonance
frequency could come from the vertical closed orbit deviations in the
quadrupole magnets of the ring. However, this contribution of $\Delta
f_r/f_r =4\times 10^{-8}$ is comparable to the in-principle
limitation of the method arising the knowledge of the deuteron
$G$-factor. It is over two orders of magnitude below the accuracy
achieved in the experiment.

%
\subsection{Determination of the deuteron beam momenta $\boldsymbol{p}$
and the momentum smearing $\boldsymbol{\Delta p/p}$}
\label{subsec:ResBeamMom}

The deuteron kinematic $\gamma$-factor and the beam momenta were
calculated according to Eq.~\eqref{eq:gamma}
\begin{eqnarray}\nonumber
    \gamma &=& \frac{1}{G_d} \left( \frac{f_r}{f_0} - 1 \right) \\
\label{eq:gamma}
    p &=& m_d \, \beta \, \gamma = m_d \; \sqrt{\gamma^2 - 1}
\end{eqnarray}
from the knowledge of the revolution and the spin-resonance
frequencies. The accuracies to which both frequencies are determined
are dominated by systematic effects. The revolution frequency
measured by the Schottky spectrum analyzer has an uncertainty of
$\Delta f_0 = 6$~Hz, corresponding to one in the beam momentum of
$50$~keV/$c$. The error in the determination of the spin resonance
frequency $\Delta f_r = 15$~Hz arises from the small variations of
the orbit length and corresponds to an uncertainty in the beam
momentum of $164$~keV/$c$. Because these systematic uncertainties are
independent, they are added quadratically to give a total uncertainty
$\Delta p/p \leqslant 6\times 10^{-5}$, i.e., a precision of
$170$~keV/$c$ for beam momenta in the range of $3100-3200$~MeV/$c$.
This is over an order of magnitude better than ever reached before
for a standard experiment in the COSY ring. An example of the
reconstructed beam properties is presented in Table~\ref{tab:typ_res}
for one typical energy setting. The measured beam momentum differed
by $\approx 5$~MeV/$c$ from the nominal requested momentum.

\begin{table}[hbt]
    \caption{\label{tab:typ_res} Typical results for one beam setting.}
    \begin{ruledtabular}
        \begin{tabular}{l | l}
            Nominal beam momentum           &       3150.5~[MeV/$c$]          \\
            \hline
            Revolution frequency                    &       $1403832\pm 6$~[Hz]     \\
            Spin-resonance frequency            &       $1011810\pm 15$~[Hz]    \\
            Orbit length                                    &       $183.4341\pm 0.0002$~[m]    \\
            Relativistic $\gamma$ factor                &       $1.9530\pm0.0001$           \\
            Reconstructed beam momentum &       $3146.41\pm \ 0.17$~[MeV/$c$]   \\
        \end{tabular}
    \end{ruledtabular}
\end{table}

Two further quantities, the beam momentum smearing $\delta p/p$ and
the smearing of the orbit length $\delta s /s$, can be extracted from
the spin-resonance spectra. As discussed in
Sec.~\ref{subsec:ResSysShift}, the measured spin-resonance widths of
80 to 100~Hz are dominated by the momentum spread. Assuming a
gaussian distribution in the revolution frequency with a
$\textrm{FWHM}=40-50$~Hz, and neglecting other effects, the width of
the spin-resonance distribution requires a momentum spread of
$\left(\delta p/p\right)_{\text{rms}} \approx 2 \times 10^{-4}$. This
upper limit on the beam momentum width corresponds to a smearing of
the orbit length of $\left(\delta s/s \right)_{\text{rms}} \approx 4
\times 10^{-5}$.

The momentum spread could be checked from the frequency slip factor
$\eta$, which was measured at each energy. Using $\delta p/p = 1/\eta
\times (\delta f_0/f_0)$, this leads for example at
$p_{\text{nominal}} = 3.1625$~GeV/$c$ to ($\delta p/p)_{\text{rms}} =
1.4 \times 10^{-4}$, which is consistent with the limit obtained from
the resonance distribution.
%
\section{Conclusions and outlook}
\label{sec:end} In this paper we have shown how to determine the
momentum of a deuteron beam in a circular accelerator with high
precision using the spin-resonance technique developed at the VEPP
accelerator for electron beams. We have studied the depolarization of
a polarized deuteron beam at COSY through an induced spin resonance
for thirteen different beam energies. This was done under standard
experimental conditions, i.e., with cavities, in particular the
\emph{bb} cavity, and a thick internal cluster-jet target. The
momenta and other beam properties were found by measuring the
position of the spin-resonance and revolution frequencies.

It was possible to determine the beam momenta with an accuracy of
$\Delta p/p\leqslant 6\times 10^{-5}$, i.e., the thirteen momenta in
the range $3100-3200$~MeV/$c$ were measured with precisions of
$\approx 170$~keV/$c$, a feat never before achieved at COSY. The
actual precision was limited by the systematic variations of the
orbit length and the characteristics of the Schottky spectrum
analyzer. The latter could be improved significantly through the
comparison with a calibrated frequency standard.

The orbit length could be extracted from the revolution and
spin-resonance frequencies with an accuracy of $\Delta s/s \leqslant
2 \times 10^{-6}$. Thus for COSY, with a circumference of 183.4~m,
the orbit length could be measured with a precision below 0.3~mm.
This may allow one to gain a better knowledge of the orbit behavior
in COSY.

These results were achieved using a deuteron
beam, but there are no in-principle reasons why the depolarization
technique should not be applicable to proton beams at COSY with same
success.

In summary, the spin-resonance method is a powerful beam diagnostic
tool for circular accelerators, synchrotrons or storage rings without
Siberian snakes to investigate and determine beam properties. In our
particular case it should eventually allow the mass of the $\eta$
meson to be measured with a precision of $\Delta m_{\eta} \leqslant
50$~keV/$c^2$.

%
\begin{acknowledgments}
The authors wish to express their thanks to the other members of the
COSY machine crew for producing such good experimental conditions and
also to the other members of the ANKE collaboration for diverse help
in the experiment. The spin-depolarizing studies for deuterons and
protons were initiated at COSY by Alan Krisch and
other members of the SPIN@COSY collaboration and we
have benefited much from their experience. This work was supported in
part by the HGF-VIQCD, and JCHP FEE.
\end{acknowledgments}

%

\end{document}